\documentclass[aps,pra,twocolumn,nopacs,floatfix]{revtex4-2}
\usepackage{epsfig}
\usepackage{graphicx}
\usepackage{dcolumn}
\usepackage{amsthm,amsmath}
\usepackage{comment}
\usepackage[colorlinks]{hyperref}
\usepackage{xcolor}
\usepackage{longtable}
\usepackage{amsmath}

\begin{document}

\title{Relativistic Normal Coupled-cluster Theory Analysis of Second- and Third-order Electric Polarizabilities of Zn I}

\author{A. Chakraborty$^{1,2}$, S. K. Rithvik$^{1,2}$ and B. K. Sahoo$^1$}

\affiliation{
$^1$Atomic, Molecular and Optical Physics Division, Physical Research Laboratory, Navrangpura, Ahmedabad 380009, India\\
 $^2$Indian Institute of Technology Gandhinagar, Palaj, Gandhinagar 382355, India
}

\begin{abstract}
We present precise values of electric polarizabilities for the ground state of Zn due to second-order dipole and quadrupole interactions, and due to third-order dipole-quadrupole interactions. These quantities are evaluated in the linear response theory framework by employing a relativistic version of the normal coupled-cluster (NCC) method. The calculated dipole polarizability value is compared with available experimental and other theoretical results including those are obtained using the ordinary coupled-cluster (CC) methods in both finite-field and expectation value evaluation approaches. We also give a term-by-term comparison of contributions from our CC and NCC calculations in order to show differences in the results from these two methods. Moreover, we present results from other lower-order methods to understand the role of electron correlation effects in the determination of the above quantities. A machine learning based scheme to generate optimized basis functions for atomic calculations is developed and applied here. From the analysis of the dipole polarizability result, accuracy of the calculated quadrupole and third-order polarizability values are ascertained, for which no experimental values are currently available.
\end{abstract}

\date{\today}

\maketitle   
    
\section{Introduction}

Atoms are spherically symmetric, but under the influence of stray electric fields, distribution of their electric charges are deformed \cite{bonin,haken}. In the ground state of a closed-shell atom, the first-order energy shift due to weak electric field vanishes and the leading order contributions to the energy shifts come from the second-order followed by third-order effects \cite{buckingham,maroulis1}. These contributions are usually factorised into powers of electric field strength and in terms of electric polarizabilities that are atomic state dependent but independent of the applied electric field strength \cite{bonin,manakov}. With the knowledge of these polarizabilities, it is possible to estimate energy shifts in an atomic system for an arbitrary weak electric field. As a result, there has been immense interest to study electric polarizabilities both experimentally and theoretically \cite{oymak,peter}. Among them, the electric dipole polarizability ($\alpha_d$) has been studied extensively due to its predominant contribution to the energy shift, followed by electric quadrupole polarizability ($\alpha_q$), while third-order polarizability ($B$) has received very little attention.

The $\alpha_d$ values of Group IIB elements Zn, Cd and Hg have been measured accurately \cite{goebel1,goebel2,goebel3} and theoretical calculations based on sophisticated many-body methods agree with the experimental values for the Zn and Hg atoms \cite{pershina,dzuba,goebel2,seth,kello,yashpal1,yashpal2,chattopadhyay,sahoo}. However, many calculations show significant deviations from the experimental values for the Cd atom \cite{seth,kello,yashpal1}. More precise knowledge of polarizabilities in these atoms is quite useful for several fundamental applications. For example, the Cd and Hg atoms from this group are being used as candidates in atomic clocks \cite{tyumenev,yamaguchi} and a precise knowledge of polarizabilities in these atoms will be required to estimate systematic effects of the atomic clocks. Hg is used to probe violation of parity and time-reversal symmetry violating interactions \cite{heckel,bks1}. As an important application, the knowledge of $\alpha_d$, $\alpha_q$ and $B$ values are needed to construct the polarization potential seen by an external electron (or positron) in the vicinity of an atom in the scattering physics problem as given by \cite{jain,tenfen}
\begin{equation} \label{eq1}
V_{\mathrm{pol}}(r)=-\frac{\alpha_{d}}{2 r^{4}}-\frac{\alpha_{q}}{2 r^{6}}+\frac{B}{2 r^{7}}-\cdots .
\end{equation} 

With the advent of modern technologies, high-precision measurements of energy shifts due to external electric fields are feasible, from which precise values of higher-order polarizabilities can be inferred. However, there are two important objectives behind pursuing theoretical studies of the electric polarizabilities of atomic systems. First, it helps demonstrate the validity of a theoretical approach by reproducing the experimental result and is able to provide insights into the behavior of electron correlation effects within the investigated atomic system. Secondly, but more importantly, the reason behind performing theoretical studies of electric polarizabilities by developing and applying sophisticated many-body methods is to provide their accurate values in systems where experimental results are not available, so as to guide future measurements of these quantities. As mentioned above, a number of calculations of $\alpha_d$ for Zn have been performed \cite{pershina,dzuba,goebel2,seth,kello,yashpal1,chattopadhyay} but there does not exist any measurement of $\alpha_q$ and $B$. Only a few theoretical studies of these quantities based on the non-relativistic formalism are available \cite{goebel2}. Owing to significant challenges, one is yet to see the experimental determination of $B$ values. It is therefore interesting to study the role of electron correlation effects by evaluating $\alpha_q$ and $B$. 

Among the commonly used atomic many-body methods for calculating spectroscopic properties, coupled-cluster (CC) theory is proven to be one of the most reliable and powerful methods \cite{bartlett,crawford,ccbook}. The CC theory has also been applied to nuclear, molecular and condensed matter systems to account for correlation effects among the subatomic particles accurately \cite{hagen,bishop,christiansen} and hence, this theory is being treated as the gold standard for many-body methods. Within the CC formalism, a variety of procedures are being proposed and implemented to evaluate various spectroscopic properties reliably \cite{bartlett,bishop}. Several calculations of atomic polarizabilities have been reported employing either the semi-empirical approaches \cite{peter,mitroy} or the less accurate numerical approaches like the finite-field method \cite{pershina,dzuba,goebel2,seth,kello}. It is possible that semi-empirical calculations can give reasonably accurate results because a part of uncertainties are excluded by making use of some experimental quantities, but they cannot demonstrate the true potency of the employed many-body methods. Due to the odd-parity nature of the dipole operator, it is not convenient to include the dipole interaction Hamiltonian in the atomic calculation to estimate energy level shifts by adopting spherical symmetry properties of atomic systems. Thus, results from the finite-field approach are estimated using programs that have been developed for evaluating molecular properties by imposing additional constraints to describe atomic systems \cite{kellman} in the Cartesian coordinate system \cite{thompson}. By employing a spherical coordinate system, the interaction Hamiltonian due to the dipole interaction can be treated perturbatively. Though the quadrupole operator is an even parity operator, it cannot be added to the atomic Hamiltonian in order to evaluate the quadrupole polarizability owing to the fact that it is a finite rank operator.

In this work, we have applied a linear response approach to determine the $\alpha_d$, $\alpha_q$ and $B$ values of Zn in the relativistic CC (RCC) theory framework by retaining the spherical symmetry properties of atoms. Earlier, we had employed this approach to determine $\alpha_d$ of Zn using the expectation value approach of the RCC theory \cite{yashpal1}. This approach involves a non-terminating series and contribution from the normalization of the wave function was not included in order to conveniently evaluate the expression by making use of the connecting terms. However, two different implementations of this method had produced quite different $\alpha_d$ results for the Cd atom \cite{yashpal1,chattopadhyay}. To avoid brute-force termination in the expression and ambiguity in accounting for the contribution from the normalization of the wave function, we have developed the relativistic normal coupled-cluster (RNCC) theory to determine the aforementioned quantities \cite{bks2}. The RNCC method has been applied previously for the accurate determination of $\alpha_d$ values for Xe, Cd and Hg atoms \cite{bks2,bks3,sakurai}, resulting in a reconciliation between the theoretical and experimental polarizability values for the Cd atom \cite{bks3}. This immediately prompted a re-analysis of the experimental data for this atom \cite{hohm}. A recent calculation of $\alpha_d$ of Cd using the finite-field approach in the RCC theory framework offers further support to these findings \cite{guo}. In view of this, it is imperative to carry out the polarizability calculations of Zn using the RNCC theory and make a comparative analysis with the calculation obtained using the expectation value approach of the RCC theory to understand better about both the methods. In this work, we have performed calculations of $\alpha_d$, $\alpha_q$ and $B$ using expectation value approach in the RCC and RNCC methods. We then compare results from both the approaches term-by-term. In addition, we give results from a lower-order method using the same basis functions in order to demonstrate the propagation of electron correlation effects at different levels of approximations in the many-body method. The results are given in atomic units (a.u.) unless otherwise stated explicitly.    

\section{Theory}

In the presence of a weak electric field $\mathcal{E}(r)$ with strength $\mathcal{E}_0$, the ground state energy level of Zn can be expressed in the perturbative approach as \cite{buckingham,goebel2,maroulis}
\begin{eqnarray}\label{eq2}
E_0 &=& E_0^{(0)} + E_0^{(1)} + E_0^{(2)} + \cdots \nonumber \\
&=& E_0^{(0)} - \frac{1}{2} \alpha_d \mathcal{E}_0^2 -\frac{1}{4} \alpha_q \left ( \frac{ \partial \mathcal{E}(r)} {\partial r} \vert_{\mathcal{E}=\mathcal{E}_0 } \right )^2  \nonumber \\
&& - \frac{1}{2} B \mathcal{E}_0 \frac{ \partial \mathcal{E}(r)} {\partial r } \vert_{\mathcal{E}=\mathcal{E}_0}  + \cdots ,
\end{eqnarray}
where $E_0^{(n)}$ denotes $n^{th}$ order correction to the energy with $E_0^{(0)}$ as the ground state energy level of the free Zn atom and the first-order energy shift to the ground state of Zn is $E_0^{(1)}=0$. When $\mathcal{E}(r)$ is generated from a charge $Q_e$ placed at a distance $r$, then the above equation is given by \cite{goebel2,tenfen}
\begin{eqnarray} \label{eq3}
E_0 &=& E_0^{(0)}- \frac{1}{2} \alpha_d \frac{ Q_e^2 } { r^4} - \frac{1}{2} \alpha_{q} \frac {Q_e^2} {r^6} + \frac{1}{4} B\frac{Q_e^3}{r^7} +\cdots .
\end{eqnarray}
On this basis, when an external charged particle like an electron or positron is seen in the vicinity of an atom, its polarization potential is constructed using Eq. (\ref{eq1}). 

With the prior knowledge of atomic wave functions $| \Psi_{k}^{(0)} \rangle$ and energies $E_{k}^{(0)}$ of the free Zn atom with $k$ representing the level of a state, we can evaluate $\alpha_d$, $\alpha_q$ and $B$ values using perturbative analysis as \cite{tenfen}
\begin{eqnarray}\label{alpha}
\alpha_d &=& -\frac{2}{ \langle \Psi_0^{(0)} \mid \Psi_0^{(0)} \rangle} \sum_{k\ne 0} \frac{|\langle \Psi_0^{(0)}|D| \Psi_k^{(0)} \rangle |^{2}}{E_0^{(0)}-E_k^{(0)}} , \\
\alpha_q &=& -\frac{2}{\langle\Psi_0^{(0)} \mid \Psi_0^{(0)}\rangle} \sum_{k \ne 0} \frac{| \langle \Psi_0^{(0)}| Q| \Psi_k^{(0)}\rangle |^{2}}{E_0^{(0)}-E_k^{(0)}}
\end{eqnarray}
and
\begin{eqnarray}
B &=& \frac{2}{ \langle \Psi_0^{(0)} \mid \Psi_0^{(0)} \rangle}  \sum_{k \neq 0} \sum_{m \neq 0} \frac{\langle \Psi^{(0)}_0|D| \Psi^{(0)}_k\rangle }{(E_0^{(0)}-E_k^{(0)}) } \nonumber \\ 
&& \times \langle \Psi_k^{(0)}| D  | \Psi_m^{(0)} \rangle  \frac{\langle \Psi_m^{(0)}|Q| \Psi_0^{(0)} \rangle}{(E_0^{(0)}-E_m^{(0)} )} ,
\end{eqnarray}
where $D$ and $Q$ are the electric dipole and quadrupole operators, respectively. Since it is impractical to evaluate the complete set of $| \Psi_{k}^{(0)} \rangle$ for the evaluation of the above quantities, they can be determined conveniently by expressing as \cite{bijaya1,bijaya2,bijaya3} 
\begin{eqnarray}
\alpha_d &=& 2 \frac{\langle \Psi_0^{(0)}|D| \Psi_0^{(d,1)} \rangle }{\langle \Psi_0^{(0)}|\Psi_0^{(0)} \rangle } , \nonumber \\
\alpha_q &=& 2 \frac{\langle \Psi_0^{(0)}|Q| \Psi_0^{(q,1)} \rangle }{\langle \Psi_0^{(0)}|\Psi_0^{(0)} \rangle }
\end{eqnarray}
and
\begin{eqnarray}
B &=& 2 \frac{\langle \Psi_0^{(d,1)}|D| \Psi_0^{(q,1)} \rangle }{\langle \Psi_0^{(0)}|\Psi_0^{(0)} \rangle } ,
\end{eqnarray}
where the first-order wave functions are defined as
\begin{eqnarray}\label{alpha1}
| \Psi_0^{(d,1)} \rangle &=&  \sum_{n\ne 0}| \Psi_n^{(0)}\rangle \frac{ \langle \Psi_0^{(0)}|D| \Psi_n^{(0)} \rangle}{E_0^{(0)}-E_n^{(0)}} 
\end{eqnarray}
and
\begin{eqnarray}
| \Psi_0^{(q,1)} \rangle &=& \sum_{n\ne 0}| \Psi_n^{(0)}\rangle \frac{ \langle \Psi_0^{(0)}|Q| \Psi_n^{(0)} \rangle }{E_0^{(0)}-E_n^{(0)}} .
\end{eqnarray}
Therefore, contributions from all the intermediate states in the sum-over-states to $\alpha_d$, $\alpha_q$ and $B$ can be accounted through the first-order wave functions by determining them as the solution of the following inhomogeneous equation
\begin{eqnarray}
(H_0 - E_0^{(0)}) |\Psi_0^{(o,1)} \rangle = - O |\Psi_0^{(0)} \rangle 
\end{eqnarray}
in the {\it ab initio} framework with the atomic Hamiltonian $H_0$ and $O$ denoting either $D$ or $Q$. As discussed in the next section, these first-order wave functions are solved in the RCC and RNCC theory approaches in this work.

\section{Methods for Calculation}

\subsection{Evaluation of properties}

In the relativistic framework, we consider the Dirac-Coulomb (DC) atomic Hamiltonian in the calculation which is given by
\begin{equation}
H_0=\sum_{i}\left[c \alpha_{i} \cdot \mathbf{p}_{i}+\left(\beta_{i}-1\right) c^{2}+V_{n u c}\left(r_{i}\right)+\sum_{j>i} \frac{1}{r_{i j}}\right] ,
\end{equation}
where $c$ is the speed of light, $\alpha$ and $\beta$ are the Dirac matrices, $V_{nuc}$ is the nuclear potential and $r_{ij}$  the inter-electronic separation between the electrons located at the $r_i$ and $r_j$ radial positions with respect to the center of the nucleus.   

We begin our calculations with the Dirac-Fock (DF) approximation ($H_0 = H_{DF} + V_{0}$ with $V_0=H_0-H_{DF}$ for the DF Hamiltonian $H_{DF}$) and obtain the exact  wave function by expressing as 
\begin{eqnarray}
|\Psi_0^{(0)} \rangle = \Omega^{(0)} |\Phi_0 \rangle ,
\end{eqnarray}
where $|\Phi_0 \rangle$ is the unperturbed DF wave function and the wave operator $\Omega^{(0)}$ accounts for the contributions from $V_0$. After including the external operator $O$, the first-order perturbed wave function can be written as
\begin{eqnarray}
|\Psi_0^{(o,1)} \rangle = \Omega^{(o,1)} |\Phi_0 \rangle .
\end{eqnarray}
In the perturbative analysis, the unperturbed and perturbed effects are accounted by expressing \cite{yashpal1,yashpal2,lindgren,yashpal3}
\begin{eqnarray}
\Omega^{(0)} = \Omega^{(0,0)} + \Omega^{(1,0)} + \Omega^{(2,0)} + \Omega^{(3,0)} + \cdots 
\end{eqnarray}
and 
\begin{eqnarray}
\Omega^{(o,1)} = \Omega^{(0,1)} + \Omega^{(1,1)} + \Omega^{(2,1)} + \Omega^{(3,1)} + \cdots , 
\end{eqnarray}
where $\Omega^{(n,m)}$ denotes inclusion of $n^{th}$ and $m^{th}$ order of $V_0$ and $O \equiv D/Q$ operators in the calculations. In the many-body perturbation theory (MBPT method), the amplitudes of these wave operators can be determined using the generalized Bloch's equation \cite{yashpal1,yashpal2,yashpal3} for each order of perturbation as given by
\begin{equation}
 \begin{aligned}
\left[\Omega^{(n,m)}, H_{DF} \right] P_0 =& Q_0 V_0 \Omega^{(n-1,m)} P_0  \\
&-  \sum_{k=1}^{n-1}  \Omega^{(n-k,m)} P_0 V_0 \Omega^{(k-1,m)} P_0, \\ 
&+Q_0 O \Omega^{(n,m-1)} P_0  \\
&-\sum_{r=1}^{m-1}  \Omega^{(n,m-r)} P_0 O \Omega^{(n,r-1)} P_0
\end{aligned}
\end{equation}
by equating terms with the same order of perturbation from both the sides, where
$P_0=| \Phi_0 \rangle \langle \Phi_0 |$ and $Q_0= 1 - P_0$. To understand how electron correlation effects propagate from the lower-order level to the higher-order level of perturbation in the determination of polarizabilities, we consider one- and two-orders of $V_0$ in the second-order (MBPT(2)) and third-order (MBPT(3)) MBPT method respectively, and estimate the $\alpha_d$ and $\alpha_q$ values. It is obvious from here that the DF values of the polarizabilities can be obtained by considering zero-order of $V_0$ in the calculation. We also intend to verify the results by approximating  $\Omega^{(0)} \approx \Omega^{(0,0)} =1$ and $\Omega^{(o,1)} \approx \sum_{n=1}^{\infty} \Omega^{(n,1)}$ but accounting only the core-polarization effects in the random-phase approximation (RPA) framework \cite{yashpal1}.

The RPA as well as all-order contributions from the non-RPA effects can be captured simultaneously by the RCC theory \cite{yashpal1,yashpal2,bijaya4}, in which the unperturbed exact wave function is given by
\begin{equation}
|\Psi_0^{(0)} \rangle=e^{T^{(0)}} |\Phi_0 \rangle , \label{eqcc1}
\end{equation}
where $T^{(0)}$ accounts for electron correlation effects from $V_0$. Analogously, the first-order perturbed wave function is given by
\begin{equation}
|\Psi_{0}^{(o,1)} \rangle = e^{T^{(0)}} T^{(o,1)} |\Phi_0 \rangle , \label{eqcc2}
\end{equation}
where $T^{(o,1)}$ includes contributions from both $V_0$ and the perturbative operator $O$. In this approach, the expressions for $\alpha_d$, $\alpha_q$ and $B$ are given by \cite{bijaya1,bijaya2,bijaya3} 
\begin{eqnarray}
\alpha_d = 2 \frac{\langle \Phi_0 | e^{T^{(0)\dagger}} D  e^{T^{(0)}} T^{(d,1)} | \Phi_0 \rangle} {\langle \Phi_0 | e^{T^{(0)\dagger}}  e^{T^{(0)}} | \Phi_0 \rangle} , \\
\alpha_q = 2 \frac{\langle \Phi_0 | e^{T^{(0)\dagger}} Q  e^{T^{(0)}} T^{(q,1)} | \Phi_0 \rangle} {\langle \Phi_0 | e^{T^{(0)\dagger}}  e^{T^{(0)}} | \Phi_0 \rangle}
\end{eqnarray}
and
\begin{eqnarray}
B = 2 \frac{\langle \Phi_0 |T^{(d,1)\dagger}e^{T^{(0)\dagger}}  D e^{T^{(0)}} T^{(q,1)}| \Phi_0 \rangle} {\langle \Phi_0 | e^{T^{(0)\dagger}}  e^{T^{(0)}} | \Phi_0 \rangle} .
\end{eqnarray}
Evaluating the above expressions involves two major challenges, even after making approximations in the level of excitations in the calculations. The first being that it has two non-terminating series in the numerator and denominator. The second that the numerator can have factors both connected and disconnected with the operators $D$ or $Q$. These present practical problems in implementing and accounting for contributions from all these terms in a convincing manner. To address these problems partially, we approach the evaluation of the $\alpha_d$, $\alpha_q$ and $B$ values in a slightly different way as described below.

Let us assume for the time being that the interaction operator $O$ is a part of the atomic Hamiltonian and given by 
\begin{eqnarray}
H = H_0 + \lambda O , \label{eqcc}
\end{eqnarray}
where $\lambda=1$ and is introduced to keep track of the order of $O$ in the calculations. The atomic wave function $| \Psi \rangle$ of the above Hamiltonian in the RCC theory can be given by
\begin{eqnarray}
|\Psi \rangle = e^{\tilde{T}} |\Phi \rangle = e^{T} |\Phi_0 \rangle, \label{eqcc3}
\end{eqnarray}
where $|\Phi \rangle$ is the modified DF wave function constructed in the presence of $O$ with the corresponding electron excitation operator $\tilde{T}$ due to both $V_0$ and $D$, while $T$ is also the electron excitation operator due to both $V_0$ and $D$ but considering excitations from $|\Phi_0 \rangle$. The expectation value of $O$ using $|\Psi \rangle$ can be mathematically given by
\begin{eqnarray}
\langle O \rangle &=& \frac{\langle \Psi | O | \Psi \rangle} {\langle \Psi | \Psi \rangle}  \nonumber \\
 &=& \frac{\langle \Phi_0 | e^{T^{\dagger}} O e^T | \Phi_0 \rangle} {\langle \Phi_0 | e^{T^{\dagger}} e^T | \Phi_0 \rangle} . \label{eqp1}
\end{eqnarray}
Following Refs. \cite{bartlett,pal}, the above expression yields
\begin{eqnarray}
\langle O \rangle &=& \langle \Phi_0 | e^{T^{\dagger}} O e^T | \Phi_0 \rangle_c , \label{eqp2}
\end{eqnarray}
where the subscript $c$ indicates that only connected terms can exist in the expression. Now expanding $T$ in powers of $\lambda$ as
\begin{eqnarray}
T = T^{(0)} + \lambda T^{(o,1)} + \mathcal{O}(\lambda^2)
\end{eqnarray}
and retaining terms linear in $\lambda$ in Eqs. (\ref{eqp1}) and (\ref{eqp2}), we get
\begin{eqnarray}
 \frac{\langle \Phi_0 | e^{T^{(0)\dagger}} O  e^{T^{(0)}} T^{(o,1)} | \Phi_0 \rangle} {\langle \Phi_0 | e^{T^{(0)\dagger}}  e^{T^{(0)}} | \Phi_0 \rangle} =  \langle \Phi_0 |  e^{T^{(0)\dagger}} O e^{T^{(0)}} T^{(o,1)} | \Phi \rangle_c . \ \ \ \ \
\end{eqnarray}
This expression is mathematically equivalent to the second-order polarizability expression. Therefore, the aforementioned polarizabilities can be evaluated in the expectation value evaluation approach of RCC theory as
\begin{eqnarray}
\alpha_d = 2 \langle \Phi_0 | e^{T^{(0)\dagger}} D  e^{T^{(0)}} T^{(d,1)} | \Phi_0 \rangle_c ,  \\
\alpha_q = 2 \langle \Phi_0 | e^{T^{(0)\dagger}} Q  e^{T^{(0)}} T^{(q,1)} | \Phi_0 \rangle_c
\end{eqnarray}
and
\begin{eqnarray}
B = & 2 \langle \Phi_0 |T^{(d,1)\dagger}e^{T^{(0)\dagger}}  D e^{T^{(0)}} T^{(q,1)}| \Phi_0 \rangle_c .
\end{eqnarray}
Though not specified explicitly, the above expression for $B$ is obtained by adding $\lambda_1 D$ and $\lambda_2 Q$ in the atomic Hamiltonian and equating terms in $\lambda_1^* \lambda_2$ and $\lambda_1 \lambda_2^*$ from the expectation value expression given by Eq. (\ref{eqp1}), where $\lambda_1$ and $\lambda_2$ are two arbitrary complex parameters. 

Before we pursue the calculations of $\alpha_d$, $\alpha_q$ and $B$ using the above connected terms in the RCC theory, we intend to point out a few issues associated with these expressions. Although it removed the non-terminating series appearing in the denominator, it still contains a non-terminating series in the numerator. Again, the above derivations of the expressions were based on the assumption that no approximation was made in $T$, but the actual calculations are carried out after approximating it to a certain level of excitations. Thus, the cancellation of the normalization of the wave function may not be exact and it will slowly tend towards exact with inclusion of higher and higher-order terms gradually. This is also true in the case when external perturbation is not included in the atomic Hamiltonian and expectation value of an operator is evaluated using Eq. (\ref{eqp2}) in the RCC theory. Nonetheless, these problems can be circumvented in the RNCC method as discussed below.

First, we want to make it clear that we shall approach in the same manner from Eq. (\ref{eqcc}) of RCC theory to derive expressions for polarizabilities in our RNCC theory. As in the usual approach of the RNCC theory, the ket state $|\Psi \rangle$ is expressed as the ordinary RCC theory but in place of $\langle \Psi |$ a new bra state $\langle \tilde{\Psi} |$ is defined for $H$ such that both $\langle \Psi |$ and $\langle \tilde{\Psi} |$ have the same eigenvalue for $H$ and it satisfies the biorthogonal condition \cite{bishop,arponen,bishop1}
\begin{eqnarray}
\langle \tilde{\Psi} | \Psi \rangle =1 .
\end{eqnarray}
Due to the fact that $\langle \Psi |$ is constructed by the de-excitation RCC operator ($T^{\dagger}$), the RNCC bra state is expressed as
\begin{eqnarray}
 \langle \tilde{\Psi} | = \langle \Phi_0 | (1+\Lambda) e^{-T} ,
\end{eqnarray}
with a de-excitation operator ${\Lambda}$. It then obviously follows that
\begin{eqnarray}
 \langle \tilde{\Psi} | \Psi \rangle = \langle \Phi_0 | (1+ \Lambda) e^{- T} e^{ T} |\Phi_0 \rangle =1 .
\end{eqnarray}
To ensure that both $\langle \Psi |$ and $\langle \tilde{\Psi}|$ have the same eigenvalue for $H$, it is imperative to impose the condition 
\begin{eqnarray}
\langle \Phi_0 |\Lambda \bar{ H} |\Phi_0 \rangle= 0 ,
\end{eqnarray}
where $\bar{H} = e^{- T} H e^{ T} = (He^T)_c$  is a terminating series. This leads to the amplitude determining equations for the $T$ and $\Lambda$ operators as
\begin{eqnarray}
\langle \Phi_0^* | \bar{H}  |\Phi_0 \rangle= 0 
\end{eqnarray}
and
\begin{eqnarray}
\langle \Phi_0 |\Lambda \bar{H} |\Phi_0^* \rangle= 0 ,
\end{eqnarray}
respectively, where $|\Phi_0^* \rangle$ is an excited state determinant with respect to $|\Phi_0 \rangle$. Now, adopting the perturbative approach of Eq. (\ref{eqcc}), we can expand
\begin{eqnarray}
\Lambda = \Lambda^{(0)} + \lambda \Lambda^{(o,1)} + \mathcal{O}(\lambda^2) .
\end{eqnarray}
When $D$ and $Q$ are included simultaneously along with parameters $\lambda_1$ and $\lambda_2$ in Eq. (\ref{eqcc}), the $\Lambda$ operator can be expanded in both $\lambda_1$ and $\lambda_2$. Consequently, the RNCC expressions for $\alpha_d$, $\alpha_q$ and $B$ can be given by
\begin{eqnarray}\label{polnrcc}
\alpha_d &=& \langle\Phi_0 |\left(1+\Lambda^{(0)}\right) \tilde{D} T^{(d,1)}+\Lambda^{(d,1)} \tilde{D} | \Phi_0 \rangle \label{eqnd} \\
\alpha_q &=& \langle\Phi_0 |\left(1+\Lambda^{(0)}\right) \tilde{Q} T^{(q,1)}+\Lambda^{(q,1)} \tilde{Q} | \Phi_0 \rangle \label{eqnq}
\end{eqnarray}
and
\begin{eqnarray}
B &=& \langle\Phi_0 |\Lambda^{(d,1)}  \tilde{D} T^{(q,1)} + \Lambda^{(q,1)}  \tilde{D} T^{(d,1)}| \Phi_0 \rangle . \label{eqnb}
\end{eqnarray}
where $\tilde{O}= (Oe^{T^{(0)}})_c$. At this juncture, we would like to mention that if the above RNCC theory derivations were made based on Eqs. (\ref{eqcc1}) and (\ref{eqcc2}) instead of starting the derivation from Eq. (\ref{eqcc3}) then we would have got extra terms like $\left(1+\Lambda^{(0)}\right)T^{(d,1)\dagger}\bar{D}$, $\left(1+\Lambda^{(0)}\right)T^{(q,1)\dagger}\bar{Q}$ and $(\Lambda^{(0)} T^{(q,1)\dagger}  \bar{D} T^{(d,1)} + \Lambda^{(0)} T^{(d,1)\dagger} \bar{D} T^{(q,1)})$ in Eqs. (\ref{eqnd}), (\ref{eqnq}) and (\ref{eqnb}) respectively. It implies that deriving the expressions for $\alpha_d$, $\alpha_q$ and $B$ from the expectation value equation given by Eq. (\ref{eqp1}) has a lot of computational advantages. Nonetheless, the final results should be independent of a theoretical approach provided an exact theory has been implemented in a consistent manner whereas in an approximated calculation, choice of implementation should be made judiciously in order to achieve more reliable results as well as a reduction in the computational cost.

It is obvious from the above expressions for the polarizabilities in the RNCC theory that they are free from  non-terminating series as appear in the RCC theory and normalization of the wave function naturally becomes one. Therefore, it is more practical to handle the calculations using the RNCC theory. It also removes the ambiguity regarding the non-exact cancellation between contributions from the normalization of the wave function and disconnected terms of the RCC theory in an approximated calculation. At present, we have considered the single and double excitations in the RCC theory (RCCSD method) and RNCC theory (RNCCSD method) by defining the electron excitation and de-excitation operators as
\begin{eqnarray}
T^{(0)} &=& T_1^{(0)} + T_2^{(0)}, \ \ \ T^{(0)\dagger} = T_1^{(0)\dagger} + T_2^{(0)\dagger}, \ \ \\
T^{(o,1)} &=& T_1^{(o,1)} + T_2^{(o,1)}, \ \ \ T^{(o,1)\dagger} = T_1^{(o,1)\dagger} + T_2^{(o,1)\dagger} \ \
\end{eqnarray}
and 
\begin{eqnarray}
\Lambda^{(0)} &=& \Lambda_1^{(0)} + \Lambda_2^{(0)}, \ \ \  \Lambda^{(o,1)} = \Lambda_1^{(o,1)} + \Lambda_2^{(o,1)} . \ \
\end{eqnarray}

Another pertinent point that we would like to mention here is that considering the next-level excitations, i.e. triple excitations, in the RCC theory will be too challenging computationally as the number of terms will be quite large. Owing to the fact that both $D$ and $Q$ are one-body operators and that there is a constraint of having connected terms in the polarizability calculations, only a limited number of additional terms will appear in the above expressions if we intend to include higher-level excitations through the RNCC theory. This is why we may anticipate a significant difference between the results from RCCSD and RNCCSD methods, while these differences can be minimized with the inclusion of higher-level excitations. Furthermore, the results will converge faster in the RNCC theory with the level of approximations compared to the RCC theory owing to the above constraint. 

To understand the differences between the results from the RCCSD and RNCCSD methods, we have given results from individual terms from both these methods and make a comparative analysis among them. It should also be noted that the MBPT(2) results are the lowest-order contribution of RPA. Therefore, we can explain the role of the lower-order and all-order core-polarization contributions to the determination of $\alpha_d$ and $\alpha_q$ by analysing the MBPT(2) and RPA results. Similarly, the MBPT(3) method introduces the lowest-order non-RPA contribution. The differences between the RPA and RCC results will represent the contributions due to non-core polarization effects to all-order and a comparison of this difference with the MBPT(3) results will give an idea about how important the non-core polarization effects are in the evaluation of $\alpha_d$ and $\alpha_q$. Since evaluation of $B$ depends on the first-order wave functions used in the determination of $\alpha_d$ and $\alpha_q$, accuracy of $B$ can be gauged from the accuracy of the calculated $\alpha_d$ and $\alpha_q$ values.

\subsection{Machine Learning based scheme for orbital optimization}

In the course of calculating accurate values of polarizabilities, it is necessary to use reliable single particle orbitals along with considering a powerful many-body method. There is a possibility that the method employed in a calculation is very accurate, but the results can still be bad due to use of poor quality single orbitals used in the construction of Slater determinants. In our approach, we need to know both the bound orbitals and continuum for pursuing the calculations. The bound orbitals can be obtained by solving differential equations, but it requires a different treatment to obtain the continuum. If two separate approaches are adopted to obtain the bound orbitals and continuum, then there can be orthogonality issues among them. Thus, we prefer to generate atomic orbitals using a single procedure by imposing orthonormality conditions among them. 

\begin{figure}[t]
    \centering
    \includegraphics[height = 7.8cm, width = 8.8cm]{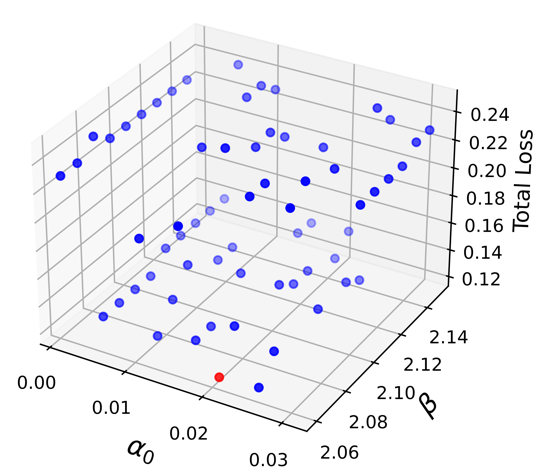}

    \caption{(color online) A plot demonstrating total loss of the single particle orbital wave functions produced using GTOs with different set of $\alpha_0$ and $\beta$ parameters with respect to numerical orbitals. The red dot corresponds to the point of least total loss which occurs at $\alpha_0 = 0.0209$  and $\beta = 2.07$.}
    \label{fig:loss}
\end{figure}

The single particle orbitals are given in the DF method as 
\begin{eqnarray} 
|\phi(r) \rangle = \begin{pmatrix} \psi_A(r)\\ \psi_B(r) \end{pmatrix}
= \frac{1}{r} \begin{pmatrix} P(r) \chi_{jl,j_z}^{P} \\ \iota Q(r) \chi_{jl,j_z}^{Q}\end{pmatrix} 
\end{eqnarray}
where $P(r)$ (or $Q(r)$) is the large (or small) component radial function, and $\chi_{jl,j_z}^{P/Q}$ is the corresponding normalized spin-angular function and is an eigenfunction of the ${\bf j}^2$, $j_z$, ${\bf l}^2$ and ${\bf s}^2$ operators. The radial functions are expressed as a linear combination of Gaussian type orbitals (GTOs) such as
\begin{equation}
P(r) = \sum_{\eta=1}^{N_b} C_\eta^L N^L g_\eta^L(r) \label{eqp}
\end{equation}
and 
\begin{equation}
Q(r) = \sum_{\eta=1}^{N_{b}} C_\eta^S N^S g_\eta^S(r)  , \label{eqq}
\end{equation}
where $C_\eta^L$ and $C_\eta^S$ are the expansion coefficients (over which the variation is performed) over a $N_b$ number of GTOs, and $N^L$ and $N^S$ are the normalization constants for the large and small components, respectively. The GTOs describing the large and small component radial functions are given by \cite{boys,mohanty}
\begin{eqnarray}
g_\eta^L(r) = r^l e^{-\alpha_\eta r^2} 
\end{eqnarray}
and
\begin{eqnarray}
g_\eta^S(r) = N^L \left (\frac{d}{dr} + \frac{\kappa}{r} \right) g_\eta^L(r) ,
\end{eqnarray}
respectively, for the relativistic quantum number $\kappa$. In Eqs. (\ref{eqp}) and (\ref{eqq}), we have unknowns as $C_\eta^L$, $C_\eta^S$, and $\alpha_\eta$. The $C_\eta^L$ and $C_\eta^S$ coefficients depend upon the choice of $\alpha_\eta$ parameters and $N_b$. Also, a suitable choice of an appropriate set of $\alpha_\eta$ values can describe the completeness of the space by a finite size of basis functions in a discretized  manner. This is also a practical requirement to carry out the calculations. Thus, it is necessary to find out only the $C_\eta^L$ and $C_\eta^S$ coefficients to describe both $P(r)$ and $Q(r)$, and consequently $|\phi(r) \rangle$. In fact, one can find optimized GTOs for molecular calculations, which make use of contracted functions to describe the large space of basis functions with minimum computational effort \cite{Dyall}. However, these functions are not suitable for describing atomic orbitals in the spherical coordinate system. In order to address this, we use reduced matrix elements for calculating atomic properties by which we simply avoid dependency of $j_z$ components of the orbitals explicitly, hence managing to include a much larger size active space in the calculations using many-body methods. Nonetheless, we solve the Roothan's equation \cite{roothan} in the relativistic framework to obtain these coefficients to construct the single particle orbitals in the DF method. To choose the $\alpha_\eta$ parameters conveniently, they are defined according to the even tempering condition \cite{bardo}, which treats $\alpha_\eta = \alpha_0 \beta^{\eta-1}$ for two arbitrary parameters $\alpha_0$ and $\beta$.  

\begin{table}[t]
    \centering
    \caption{Calculated static dipole ($\alpha_d$) and quadrupole ($\alpha_q$) polarizability values (in a.u.) of Zn using different many-body methods in the relativistic framework.}
    \begin{tabular}{p{2.5cm}p{0.3cm}  p{1.5cm} p{1.5cm} p{1.5cm}}
    \hline \hline
   Method   & & $\alpha_d$ & $\alpha_q$ & $B$ \\
         \hline
          & & & \\
      DF &  & 37.29 & 278.69  & $-1576.86$\\
      MBPT(2) & & 43.50  &376.62 & \\
      MBPT(3) & & 38.68 &340.85 & \\
      RPA & & 50.81  & 431.75 & \\
      RCCSD & & 40.32 & 318.79 & $-2809.53$ \\
      RNCCSD & & 38.99 & 314.40 & $-2195.15$\\
      \hline \hline
    \end{tabular}
    \label{tab1}
\end{table}

It is a challenge to search for a suitable set of $\alpha_0$, $\beta$ and $N_b$ that can aptly describe the DF orbitals in atomic systems. Finding an optimized set of basis functions to describe single particle orbitals can minimize the error due to the finitude of the basis set chosen and thereby improve the accuracy of the calculations. It is, however, not possible to do so by choosing the above parameters manually. Therefore, it is pertinent to find a scheme wherein we can find the optimal values for these parameters such that they can produce the DF orbitals with a high quality, meaning that the properties calculated from them should be physically more meaningful. In order to achieve this goal, we borrow the concept of a Loss Function $L$ \cite{DeepLearning} which is an essential ingredient in optimization schemes used in \textit{Machine Learning} and elsewhere. This necessitates the need for a reference data (i.e. for natural orbitals) with which we can compare our candidate data. In our case, the reference data are taken from the numerical solutions of the DF equations as solved by the GRASP-2K package \cite{joensson}. We then choose the mean squared error (MSE) loss which provides a quantitative estimate of the closeness of our candidate data to the reference data. The MSE loss is given by \cite{DeepLearning,nik}
\begin{eqnarray}
L = \frac{1}{N_t}\sum_{i}^{N_t}(y_i - \hat{y}_i)^2 ,
\end{eqnarray} 
where $N_t$ is the total size of the basis set, $y_i$ are the reference values and $\hat{y}_i$ are the candidate values. In our case, the reference data consists of the large and small components of the bound DF orbitals of the considered atom.

\section{Results \& Discussion}

In order to obtain the single particle orbitals using the optimised $\alpha_0$ and $\beta$ values, we compute the net MSE loss as a sum of individual losses due to the large ($L_{large}$) and small ($L_{small}$) components of radial function as $L = W_L L_{large} + W_S L_{small}$ with the respective weight factors $W_L$ and $W_S$. Since accurate calculations of polarizabilities mostly depend on the large component of the single particle wave function, we take a larger value for $W_L$ than $W_S$. Again, optimizing the smaller radial components are very sensitive to numerical accuracy owing to their drastically smaller magnitudes. From this point of view, we consider $L=0.8L_{large} + 0.2L_{small}$, and the initial values for the parameters are chosen as $\alpha_0 = 0.0009$ and $\beta = 2.15$. We perform a grid-search for the local optima for a given range and step size in a region of interest. We show a plot showing the MSE loss for various values of $\alpha_0$ and $\beta$ values in Fig. \ref{fig:loss}. This gives us the local optimal values for the basis parameters as $\alpha_0 = 0.0209$  and $\beta = 2.07$, using which we have performed the rest of the calculations. 

\begin{figure}[t]
    \centering
    \includegraphics[width=8.5cm,height=7cm]{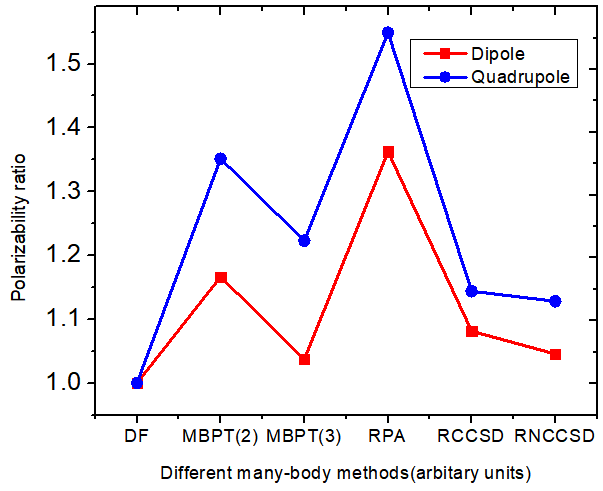}
    \caption{(color online) Ratios of dipole and quadrupole polarizability values from different many-body methods and their DF values.}
    \label{fig2}
\end{figure}

In Table \ref{tab1}, we present the results for $\alpha_d$ and $\alpha_q$ from our RNCCSD method along with results from our DF, MBPT(2), MBPT(3), RPA and RCCSD methods. If we compare the results from all these methods apart from the RNCCSD method with that are reported in Ref. \cite{yashpal1}, we find that the results are slightly different at the given level of approximation.
In Ref. \cite{yashpal1} contributions from partial triple excitations were considered in the RCCSD calculation (CCSD$_p$T) method, but the main reason for discrepancy in the result from the present work is owing to the use of optimized basis functions. A slightly large difference is seen at the MBPT(3) method because of consideration of contributions from a few additional Goldstone diagrams here which were not included in Ref. \cite{yashpal1}. The RNCCSD results are given for the first time. The trends of the $\alpha_d$ and $\alpha_q$ values through different many-body methods are shown pictorially in Fig. \ref{fig2} after normalizing with their respective DF values. As can be seen from the above table and from the figure, the difference between the DF and RNCCSD values for $\alpha_d$ is small but there is a relatively large difference for the $\alpha_q$ value from these methods. This implies that the electron correlation trends between both the properties are different. In fact, results from the lower-order methods show that the $\alpha_d$ and $\alpha_q$ values increase in the MBPT(2) method compared to the DF values, while their magnitudes reduce slightly in the MBPT(3) method. As it was mentioned earlier, the MBPT(2) method contains the lowest-order RPA correlation terms while the MBPT(3) method introduces the lowest-order non-RPA correlation terms. Thus, the above trends in the results from the MBPT(2) and MBPT(3) methods suggest that there are cancellations in the correlation contributions arising through the RPA and non-RPA types of correlations. This is further evident from the RPA and RCCSD results. The RPA, which accounts for correlation contributions due to all-order core-polarization effects, give very large values for both $\alpha_d$ and $\alpha_q$. However, the RCCSD results are close to the MBPT(3) values. This means that the core-polarization effects enhance the magnitudes of both the polarizabilities while other correlation effects contribute with opposite signs at the all-order perturbation level. We also observe that these cancellations are slightly larger (in percentage) for $\alpha_q$ than $\alpha_d$. Comparing results from the RCCSD and RNCCSD methods, the differences in $\alpha_d$ and $\alpha_q$ from both the methods are found to be about 3\% and 1\% respectively. In the above table, we have also given $B$ values but only from the DF, RCCSD and RNCCSD methods. The reason for not giving results from the other lower methods is that the theoretical evaluation of $B$ depends on the first-order wave functions due to $D$ and $Q$, so by analysing results for $\alpha_d$ and $\alpha_q$ using lower-order methods propagation of electron correlation effects from lower-order to all-order methods can be understood. 

\begin{table}[t]
\centering
\caption{Comparison of contributions to $\alpha_d$ (in a.u.) from various terms of the RCCSD and RNCCSD methods. We also compare the corresponding contributions from previously reported calculations using the RCCSD method. Here, Norm represents the difference between the contributions after
 and before normalizing the wave function with a normalization factor, NA stands for not applicable and Nonlin corresponds to the contributions coming from nonlinear terms. These abbreviations are followed in the remainder of this work.}
\begin{tabular}{p{1.6cm} p{1.1cm} p{1.1cm} p{1.1cm} p{0.1cm} p{1.6cm} p{1.1cm}}
    \hline\hline
     \multicolumn{4}{c}{RCC method} & & \multicolumn{2}{c}{RNCC method}  \\
     \cline{1-4} \cline{6-7} \\
     Term & Ref. \cite{chattopadhyay} & Ref. \cite{yashpal1} & Ours & & Term & Ours \\ 
    \hline \\
   $ DT_1^{(1)}$ & 22.795 & 21.906 & 23.793  & & $DT_1^{(1)}$ & 23.793\\
   $T_1^{(1)^\dagger}D$& 22.795 & 21.906  & 23.793 &  & $\Lambda_1^{(1)}D$ & 18.510\\
$T_1^{(0)^\dagger}DT_1^{(1)}$ & $-0.951$ & $-1.229$ & $-1.336$ & & $\Lambda_1^{(0)}DT_1^{(1)}$ & $-0.664$  \\
$T_1^{(1)^\dagger}DT_1^{(0)}$ & $-0.951$ & $-1.229$ & $-1.336$ & & $\Lambda_1^{(1)}DT_1^{(0)}$ & $-1.038$ \\
   $T_1^{(1)^\dagger}DT_2^{(0)}$& $-0.925$ & $-2.643$ & $-2.794$  &  & $\Lambda_1^{(1)}DT_2^{(0)}$ & $-2.159$ \\
      $T_2^{(0)^\dagger}DT_1^{(1)}$& $-0.925$ & $-2.643$ & $-2.794$  & &  $\Lambda_2^{(0)}DT_1^{(1)}$ & 0.0 \\
$T_1^{(0)^\dagger}DT_2^{(1)}$& 0.041 & NA &  0.072 &  & $\Lambda_1^{(0)}DT_2^{(1)}$ & 0.042 \\ 
$T_2^{(1)^\dagger}DT_1^{(0)}$& $0.041$ & NA & 0.072   &  & $\Lambda_2^{(1)}DT_1^{(0)}$ & 0.0 \\
   $T_2^{(0)^\dagger}DT_2^{(1)}$& 0.673 & 1.024 & 1.025 & & $\Lambda_2^{(0)}DT_2^{(1)}$  & 0.980 \\
   $T_2^{(1)^\dagger}DT_2^{(0)}$& 0.673 & 1.024 & 1.025 & & $\Lambda_2^{(1)}DT_2^{(0)}$  &  0.899\\
   Nonlin & NA & 0.551 & $-1.200$ &  &Nonlin   & $-1.373$ \\
   Norm  & $-4.086$ & 0.0 &  0.0 &  &  Norm & 0.0 \\
   \hline  \hline
    \end{tabular}
    \label{tab2}
\end{table}

In order to understand the differences between the RNCCSD and RCCSD values better, we give results from individual terms of these methods for $\alpha_d$, $\alpha_q$ and $B$ in Tables \ref{tab2}, \ref{tab3} and \ref{tab4}, respectively. In Table \ref{tab2}, we also give the corresponding contributions to $\alpha_d$ from the RCCSD method that were reported previously in Refs. \cite{yashpal1,chattopadhyay}. It can be seen from the term-by-term comparison between the RCCSD and RNCCSD results from the present calculations that contributions arising through the complex conjugate (c.c.) terms and the counter RNCC terms are quite different. Comparing values of individual RCC terms among the earlier calculations \cite{yashpal1,chattopadhyay} and ours, we find the trends from different terms differ. There is a similarity in the trends between the present work and Ref. \cite{yashpal1} as the implementation procedures of the RCC method is same in these calculations but the basis functions used in both the cases are different. We have used a much larger basis set of functions with 40, 39, 38, 37, 36, and 35 GTOs for the $s$, $p$, $d$, $f$, $g$ and $h$ orbitals respectively, whereas only 35 GTOs were used for each symmetry up to $g$ orbitals in Ref. \cite{yashpal1}. Furthermore, we have optimized the the GTO parameters by adopting a Machine Learning based optimization technique this time as mentioned earlier. It should be noted that calculations in Ref. \cite{chattopadhyay} and in the present work are carried out using orbitals up to $h$-symmetry and same levels of approximations are considered in the RCC theory, but the implementation procedures are different in these works. Now comparing the correlation trends through the individual RCC terms of Ref. \cite{chattopadhyay} with our calculation, we find the difference in the result from $DT_1^{(1)}$ (along with its c.c. term) is small but they differ substantially among other RCC terms. We notice that contribution arising through the normalization of the wave function (quoted as `Norm' in Table \ref{tab2}) in Ref. \cite{chattopadhyay} is quite large. In fact, it is larger than the difference between our DF and final RCC results (i.e. the net correlation contributions). In this view, we feel that by implementing the RCC theory in which only the connected terms are retained (also Norm factor does not appear) in Eq. (\ref{eqp1}) is more credible. Nonetheless, our RNCC theory takes care of this normalization factor in a natural manner. Unlike $\alpha_d$ calculations, $\alpha_q$ value of Zn was not evaluated using the linear response RCC theory earlier. Therefore, we could not make a comparative study between the contributions from our RCC terms with any earlier study in Table \ref{tab3} but show only the comparison of contributions from various terms of the RCCSD and RNCCSD methods.

\begin{table}[t]
    \centering
    \caption{Comparison of contributions to $\alpha_q$ (in a.u.) from various terms of the RCCSD and RNCCSD methods.}
    \begin{tabular}{p{2.0cm} p{1.5cm} p{0.3cm} p{2.0cm} p{1.5cm}} 
    \hline\hline
     \multicolumn{2}{c}{RCC method} &  &\multicolumn{2}{c}{RNCC method}  \\
     \cline{1-2} \cline{4-5} \\
     Term &  Results & & Term & Results \\  
    \hline \\
    $QT_1^{(1)}$& 179.704  &  & $QT_1^{(1)}$ & 179.704\\
    $T_1^{(1)^\dagger}Q$& 179.704  &  & $\Lambda_1^{(1)}Q$ & 154.456\\
    $T_1^{(0)^\dagger}QT_1^{(1)}$  & $-22.338$ & & $\Lambda_1^{(0)}QT_1^{(1)}$ & $-10.597$ \\
    $T_1^{(1)^\dagger}QT_1^{(0)}$ & $-22.338$ & & $\Lambda_1^{(1)}QT_1^{(0)}$ & $-19.628$ \\
    $T_1^{(1)^\dagger}QT_2^{(0)}$& $-1.228$  &  & $\Lambda_1^{(1)}QT_2^{(0)}$ & $-1.045$ \\
    
     $T_2^{(0)^\dagger}QT_1^{(1)}$&  $-1.228$ &  & $\Lambda_2^{(0)}QT_1^{(1)}$ &  0.0 \\
     
     $T_1^{(0)^\dagger}QT_2^{(1)}$ & 0.474 &   & $\Lambda_1^{(0)}QT_2^{(1)}$ & 0.277 \\
     
$T_2^{(1)^\dagger}QT_1^{(0)}$ & 0.474    &  & $\Lambda_2^{(1)}QT_1^{(0)}$ & 0.0 \\
     
    $T_2^{(0)^\dagger}QT_2^{(1)}$& 10.520 &  & $\Lambda_2^{(0)}QT_2^{(1)}$  & 10.131 \\
    $T_2^{(1)^\dagger}QT_2^{(0)}$& 10.520 & & $\Lambda_2^{(1)}QT_2^{(0)}$  &  9.641\\
    Nonlin & $-15.474$  & & Nonlin   &  $-8.539$ \\
    Norm &  0.0   &    & Norm & 0.0\\
     \hline  \hline
    \end{tabular}
    \label{tab3}
\end{table}

As was mentioned earlier, the ket state in the RNCC theory is the same as the one in RCC theory. Thus, the differences in results between the RCCSD and RNCCSD methods are due to different contributions arising through various de-excitation operators in both the methods. From the comparison of contributions from individual terms in the above table it is clear that the amplitudes of the de-excitation operators in the RNCCSD method are lower in magnitudes than their corresponding operators in the RCCSD method. These trends from individual terms are almost similar during the evaluation of both $\alpha_d$ and $\alpha_q$. We also find that certain terms which contribute finitely in the RCCSD method, do not contribute in the RNCCSD method as they cannot give rise connected Goldstone diagrams. This is how the lower contributions arising through the de-excitation operators of the RNCCSD method get compensated with the contributions from the extra terms of the RCCSD method. 

\begin{table}[t]
\caption{Comparison of contributions to $B$ (in a.u.) from
various terms of the RCCSD and RNCCSD methods.}
    \centering
    \begin{tabular}{p{2.0cm} p{1.5cm} p{0.3cm} p{2.0cm} p{1.5cm}} 
    \hline\hline
     \multicolumn{2}{c}{RCC method} &  &\multicolumn{2}{c}{RNCC method}  \\
     \cline{1-2} \cline{4-5} \\
     Term &  Results & & Term & Results \\ 
     \hline \\  
     $T_1^{(d,1)^\dagger} DT_1^{(q,1)}$  & $-1267.13$ & & $\Lambda_1^{(d,1)} DT_1^{(q,1)}$ &  $-987.22 $   \\
     
      $T_1^{(q,1)^\dagger}DT_1^{(d,1)}$ &  $-1267.13$ & &$\Lambda_1^{(q,1)}DT_1^{(d,1)}$ & $ -1091.26 $ \\
      
       $T_1^{(d,1)^\dagger}DT_2^{(q,1)}$ &  $-269.67$ & & $\Lambda_1^{(d,1)}DT_2^{(q,1)}$ & $-206.59$ \\
       
        $T_2^{(q,1)^\dagger}DT_1^{(d,1)}$ &  $-269.67$ & &$\Lambda_2^{(q,1)}DT_1^{(d,1)}$  & 0.0  \\
        
         $T_2^{(d,1)^\dagger}DT_1^{(q,1)}$ &  $-107.99$ & &$\Lambda_1^{(q,1)}DT_2^{(d,1)}$  & $-93.05$  \\
         
          $T_1^{(q,1)^\dagger}DT_2^{(d,1)}$ &  $-107.99$ & & $\Lambda_2^{(d,1)}DT_1^{(q,1)}$  &  0.0\\
          
           $T_2^{(q,1)^\dagger}DT_2^{(d,1)}$ &  145.63 & & $\Lambda_2^{(q,1)}DT_2^{(d,1)}$ &  20.81\\
         
          $T_2^{(d,1)^\dagger}DT_2^{(q,1)}$ &  145.63 & &$\Lambda_2^{(d,1)}DT_2^{(q,1)}$ & 127.73 \\
          
          Nonlin & 188.79  & & Nonlin & 34.43\\
      
      \hline \hline
    \end{tabular}
   \label{tab4}
\end{table}

We now turn to presenting the $B$ values from the RCCSD and RNCCSD methods. Compared to $\alpha_d$ and $\alpha_q$, very few theoretical studies of $B$ have been carried out in  atomic systems and mostly they have been reported using the FF approach. Inferring their experimental results are extremely challenging, thus accurate evaluation of $B$ is quite interesting to understand roles of various correlation effects associated in its evaluation. Using the FF approach it is possible to achieve the final value of $B$ at a given level of approximation in the many-body method, however, a linear response approach could demonstrate underlying roles of different electron correlation effects in the calculation of $B$ through various physical interactions explicitly. Since $B$ is evaluated using the first-order perturbed wave functions that are used to estimate the $\alpha_d$ and $\alpha_q$ values, the electron correlation trends of $B$ can be somewhat guessed from the earlier analyses of $\alpha_d$ and $\alpha_q$ results at different levels of approximations in the many-body methods, but the additional inter-correlation among the dipole and quadrupole operators in the evaluation of $B$ may offer quite a different picture. To fathom this, we make a comparative analysis of individual contributions to $B$ values from individual terms of the RCCSD and RNCCSD methods in Table \ref{tab4}. As can be seen from this table the most dominant contribution to $B$ comes from $T_1^{(d,1)^\dagger} DT_1^{(q,1)}$, followed by $T_1^{(d,1)^\dagger}DT_2^{(q,1)}$ then from  $T_2^{(d,1)^\dagger}DT_2^{(q,1)}$ (along with their c.c. terms) in the RCCSD method. Comparing contributions from  $T_1^{(d,1)^\dagger}DT_2^{(q,1)}$ and $T_1^{(q,1)^\dagger}DT_2^{(d,1)}$, we understand that the amplitudes of the perturbed single excitation RCC operator due to dipole operator dominate over the perturbed single excitation RCC operator due to quadrupole operator. It can be noticed from the above table that contributions from $T_2^{(q,1)^\dagger}DT_2^{(d,1)}$ and its c.c. term have an opposite sign compared to other terms. In the RNCCSD method, the contributions from the counterparts of the RCCSD method follow similar trends. As can be noticed, all terms of the RNCCSD method are distinctly different from those of the RCCSD method in contrast to the cases of $\alpha_d$ and $\alpha_q$, where half of the RNCCSD contributions were arising from the RCCSD terms. Analogous to the determination of $\alpha_d$ and $\alpha_q$, we also find that there are several terms which do not contribute to the $B$ value in the RNCCSD method whose counterpart terms in the RCCSD method do give finite contributions. This is owing to the terminating series that appear in the expression for the evaluation of $B$ in the RNCCSD method against the non-terminating series of the RCCSD method. Also, it is seen that contribution from  the $\Lambda_1^{(d,1)^\dagger}DT_1^{(q,1)}$ term is lower than the contribution from the  $\Lambda_1^{(q,1)^\dagger}DT_1^{(d,1)}$ in the RNCCSD method. This may be owing to larger magnitude of the single excitation RCC operator due to the dipole operator.

\begin{table}[t]
\caption{List of recommended values for $\alpha_d$, $\alpha_q$ and $B$ (in a.u.) from different calculations using sophisticated many-body methods, experiments and scaling procedures. Our recommended values are shown in bold font.}    \label{tab5}
\begin{tabular}{l c  c c }
\hline
\hline
Quantity &   Value & Method & Reference\\
        \hline \\
$\alpha_d$  &  \bf 38.99(31)  &  RNCCSD & This work \\
       &  40.32   &  RCCSD & This work\\
       &  38.8(8) & Expt.$+$extrapol. & \cite{goebel2} \\
       & 42.79 & CCSD & \cite{goebel2} \\
       & 41.69 & CCSD(T) & \cite{goebel2} \\
       & 39.2(8) & CCSD(T)$+$scaling & \cite{goebel2} \\
       & 41.83  & CCSD & \cite{seth} \\
       & 39.27 & RCCSD & \cite{seth} \\
       & 40.55 & CCSD(T) & \cite{seth} \\
       & 38.01 & RCCSD(T) & \cite{seth} \\
       & 41.6 & CCSD & \cite{kello} \\
       & 39.02 & CCSD$^*$ &  \cite{kello} \\
       & 40.39 & CCSD(T) &  \cite{kello} \\
       &  37.86 & CCSD(T)$^*$  & \cite{kello} \\
       & 38.666(96) & CCSD$_p$T & \cite{yashpal1} \\
       & 38.72 & PRCC & \cite{chattopadhyay} \\
       & 38.76 & PRCC(T) & \cite{chattopadhyay} \\
       &  38.92 & Expt.$+$fitting & \cite{qiao2012} \\
       & 35.33 & MCDF$+$scaling & \cite{bindiya} \\
\hline \\
$\alpha_q$ & \bf 314(4)  &  RNCCSD &  This work \\

      & 318.79   & RCCSD  & This work\\
      
       & 360.16 & CCSD  & \cite{goebel2} \\

      & 351.24 & CCSD(T)  & \cite{goebel2} \\

      & 324.8(16.2)  & CCSD(T)$+$scaling & \cite{goebel2} \\
        
\hline \\
$B$   & $\bf -2195(50)$  & RNCCSD & This work \\

   & $-2809.53$ & RCCSD & This work\\
   & $-2940$ & CCSD &  \cite{goebel2} \\
   & $-2780$ & CCSD(T) &  \cite{goebel2} \\
   & $-2370(240)$  & CCSD(T)$+$scaling & \cite{goebel2} \\
        \hline \hline 
        \end{tabular} \\
$^*$After considering quasi-relativistic corrections. \\
\end{table}

Our final recommended values from the RNCCSD method along with the previously calculated and experimental results for $\alpha_d$, $\alpha_q$ and $B$ are given in Table \ref{tab5}. The estimated uncertainties from our calculations are quoted beside the recommended values. These uncertainties account for the extrapolated contributions from the high-lying basis functions that are not included in the many-body methods, the neglected contributions from the higher-order Breit and QED interactions and from the neglected higher-level excitations. Errors due to the extrapolated basis functions and relativistic effects are analysed by employing the MBPT and RPA methods, while uncertainties due to the higher-level excitations are estimated by analysing contributions from the dominant triple excitations in the perturbative approach. Individual contributions from various sources to the above quantities are listed in Table \ref{tab6} and the net uncertainty to the final value is given by adding all the contributions in quadrature. From these analyses, we recommend the values for $\alpha_d$ to be 38.99(31) a.u, for $\alpha_q$ to be 314(4) a.u and for $B$ to be $-2195(50)$ a.u.. We have also listed the previously reported experimental and calculated values of $\alpha_d$, $\alpha_q$ and $B$ in the above table. Results from the CC and RCC methods are given from both the CCSD method and the CCSD method with contributions from partial triple excitations (CCSD(T) method) along with their relativistic versions. The differences in the results from the CCSD and CCSD(T) methods can indicate importance of the neglected contributions from the triple excitations. The experimental value of $\alpha_d$ listed in Table \ref{tab5} was measured by using Michelson twin interferometer technique \cite{goebel2}. Later, this value has been revised by fitting the data using better numerical analyses \cite{qiao2012}. Our recommended value from the RNCCSD method agrees quite well with both the values. The latest calculation of $\alpha_d$ employs a sum-over-states approach, by combining only a few E1 matrix elements from the multiconfiguration Dirac-Fock (MCDF) method, experimental energies and the rest of the contributions are estimated using lower-order methods. This shows poor agreement with the experimental result \cite{bindiya}. The calculations reported in Refs. \cite{yashpal1, chattopadhyay} are equivalent to our RCCSD method, while calculations reported in Refs. \cite{kello, seth, goebel2} are based on the FF approach using non-relativistic CC method. There seems to be an overall good agreement among all these calculations. It is worth mentioning that the recommended values given by the non-relativistic CC calculations of $\alpha_d$ in Refs. \cite{kello, seth, goebel2} have used scaled values in order to quote more accurate values, but the actual calculations give very different values. Nonetheless, agreement of the previously reported accurate values of $\alpha_d$ for Zn with our RNCCSD result suggests that our estimated $\alpha_q$ and $B$ values using this method are reliable. There is one more calculation of $\alpha_q$ and $B$ reported using the non-relativistic CC method in the FF approach \cite{goebel2}. As mentioned above, the final recommended values from these references are scaled values and the actual calculated values are again very different. For example, the CCSD results for $\alpha_q$ and $B$ are quoted in the above reference as $360.16$ a.u. and $-2940$ a.u. respectively. In the CCSD(T) method, the $\alpha_q$ value was modified to $351.24$ a.u. while the $B$ value was modified to  $-2780$ a.u. \cite{goebel2}. After scaling the result, the recommended values for $\alpha_q$ and $B$ were given as $324.8(16.2)$ a.u. and $-2370(240)$ a.u. respectively. Our RNCCSD results are obtained in the {\it ab initio} framework and they account for relativistic effects. It is, however, interesting to see that our results validate the recommended values that were reported in Ref. \cite{goebel2}. It can be further noted that our RCCSD value for $B$ given in Table \ref{tab1} differs substantially from the RNCCSD result. Therefore, good agreements of our RNCCSD results with the previously recommended values of $\alpha_d$, $\alpha_q$ and $B$ suggests that approximated RNCC method is more reliable compared to the approximated RCC method for the determination of the above quantities.

\begin{table}[t]
\caption{Estimated uncertainties to $\alpha_d$, $\alpha_q$ and $B$ values (in a.u.) from basis extrapolation (given as ``Basis"), neglected Breit interaction (quited ``Breit") and lower-order QED corrections (given as ``QED") are listed.}    \label{tab6}
\begin{tabular}{p{2.5cm}p{1.5cm} p{1.5cm} p{1.5cm}}
    \hline \hline
Source & $\alpha_d$ & $\alpha_q$ & $B$ \\
        \hline \\
Basis  &  0.11 & 0.5 & 3 \\   
Breit   & 0.05 & 0.8 & 8  \\
QED     & 0.03 & 0.4 & 4  \\
Triples & 0.30 & 4.0 & 49\\
        \hline \hline 
        \end{tabular} \\
\end{table}

\section{Conclusion}

We have employed normal coupled-cluster theory in the relativistic framework to determine the dipole, quadrupole and dipole-quadrupole interaction polarizabilities of the zinc atom. By considering the singles and doubles excitation approximation and estimating uncertainties from the neglected contributions, very accurate values for the above quantities are reported. We have also given values from other methods including ordinary relativistic coupled-cluster theory from our calculations. It was found that the dipole and quadrupole polarizabilities from the normal and ordinary coupled-cluster theory agreed quite well, but there was a large difference between the dipole-quadrupole interaction polarizability. We also compared our results with the earlier recommended values from various calculations and observe that our results from the relativistic normal coupled-cluster theory match better with those values than the results obtained using the ordinary coupled-cluster theory at the singles and doubles approximation.

\section*{Acknowledgement}

The computations reported in the present work were carried out using the Vikram-100 HPC cluster of the Physical Research Laboratory (PRL), Ahmedabad, Gujarat, India.

\end{document}